\RequirePackage{fix-cm}
\documentclass[smallextended]{svjour3}       
\smartqed  
\usepackage{graphicx}
\usepackage[numbers]{natbib}

\begin{document}

\title{Design Evaluation of Serial and Parallel sub-mK Continuous Nuclear Demagnetization Refrigerators
}


\author{David Schmoranzer\and
James Butterworth\and
S\'{e}bastien Triqueneaux\and
Eddy Collin\and
Andrew Fefferman
}


\institute{D. Schmoranzer \at
							Charles University, Ke Karlovu 3, 121 16, Prague, Czech Republic\\
              \email{david@superfluid.cz}\\
					\and J. Butterworth \and S. Triqueneaux \and E. Collin \and A. Fefferman \at
							Univ. Grenoble Alpes, CNRS, Grenoble INP, Institut Néel, 38000 Grenoble, France	
}

\date{Received: date / Accepted: date}

\maketitle

\begin{abstract}
We present the evaluation of two different design configurations of a two-stage PrNi$_5$ continuous nuclear demagnetization refrigerator. Serial and parallel configurations of the two stages are considered, with emphasis on the attainable cooling power at sub-mK temperatures and the impact of the design choices on the operation of the refrigerator. Numerical simulations of heat transfer in the setup are used to evaluate the performance of the refrigerator as well as the technological requirements for the essential thermal links. In accord with similar findings for adiabatic demagnetization refrigerators [Shirron, \emph{Cryogenics} \textbf{62}, 2014], our simulations show that the performance of both configurations improves as the thermal links improve, and that the parallel configuration yields a higher cooling power than the series design for a given thermal link resistance and sample temperature.

\keywords{ultra low temperatures \and adiabatic nuclear demagnetization \and continuous refrigeration techniques \and PrNi$_5$}

\end{abstract}

\section{Introduction}
\label{intro}
Recently, significant effort was devoted to the development of a novel ultra-low temperature refrigerator ~\cite{Fukuyama,CNDR} which would be able to overcome the limitations of traditional nuclear demagnetization setups~\cite{Lounasmaa,Pobell} and continually maintain sub-mK temperatures. Such a continuous nuclear demagnetization refrigerator (CNDR) would find use in ultra-low temperature research of numerous physical phenomena such as superfluidity of $^3$He~\cite{3HeHQV,3HeMF,3HeVIS}, quantum states of macroscopic objects~\cite{Quantum1,Quantum2}, or dissipation in amorphous matter~\cite{AmorphousSi,am1,am2,DPO} or modern materials such as graphene~\cite{Graphene}. The main factor limiting the use of traditional nuclear demagnetization setups is the fact that at sub-mK temperatures, the duration of the experimental window is often less than the thermal relaxation times of the materials of interest~\cite{Pobell} (spin-lattice relaxation in non-conducting materials, relaxation processes in amorphous materials, heat exchange between metal walls and liquid helium), which precludes proper sample thermalization. This issue becomes crucial if the demagnetization setup is mounted on a dry dilution refrigerator, as these typically have higher intrinsic heat leaks due to vibrations from the pulse tube assembly~\cite{dry}, but even on ``wet'' systems, the thermal cycling might be disruptive for precise measurements. Additionally, the ability to maintain sub-mK temperature indefinitely may provide an important stepping stone for the development of further cooling techniques and eventually lead to the experimental realization of novel states of matter, such as the dual Bose-/Fermi- type superfluidity in $^3$He-$^4$He mixtures~\cite{Tuoriniemi}.

Previously, similar designs of a continuous (electronic) adiabatic demagnetization refrigerator (CADR) were discussed in detail in the literature~\cite{Shirron0,Shirron1}, and a review may be found in Ref.~\cite{Shirron2}. The design presented below builds on the above-mentioned work and extends the temperature range of the refrigerator to sub-mK temperatures by replacing paramagnetic salt pills with a suitable nuclear magnetic refrigerant, in our case PrNi$_5$. The fundamental principle of operation of the CNDR is the same as for CADR, except that it is the entropy of nuclear spins rather than of outer-shell electronic spins that is manipulated using the external magnetic field.

The construction of the CNDR requires the use of at least two nuclear demagnetization stages with separate superconducting solenoids providing the magnetic field. The stages have to be connected to each other and/or to the sample space via thermal links consisting of metal wires/blocks and superconducting heat switches. Gas-gap heat switches are unavailable in the relavant temperature range (below 10~mK) as they become inefficient at temperatures below $\approx$250~mK~\cite{Shirron1}. In previous work on CNDR~\cite{Fukuyama,CNDR} it was shown that thermal links are indeed the crucial elements of the entire setup, requiring the use of highest purity metals prepared using specialized treatments and contacting techniques. Specifically, it was demonstrated by means of numerical simulations~\cite{Fukuyama,CNDR} that the equivalent electrical resistance (from the Wiedemann-Franz law) of the thermal link between the two demagnetization stages \emph{in a series configuration} should be comparable to 150~n${\rm \Omega}$ or less (corresponding to a thermal conductance of $1.6 \times 10^{-4}$~WK$^{-1}$ at 1 mK) to attain a cooling power of 20~nW at 1~mK, see also Fig.~\ref{fig:comparison} further below.

Practically, to obtain such a low resistance, one would need to use 5N or 6N purity aluminium blocks (RRR over 5000 was obtained in CNRS Grenoble on 6N aluminium), 5N or better copper wires annealed under oxygen (RRR over 10000 was obtained) or silver wires of equal purity as well as specialized contacting techniques (stable Al to Cu contacts were made with contact resistances of 30~n${\rm \Omega}$ or less on the area of several cm$^2$). Another crucial issue is the contact to PrNi$_5$ and the purity of the nuclear refrigerant (RRR values near 30 have been reported~\cite{Kubota}). The technical developments and the details of processes used in the construction of the CNDR setup represent a separate topic and will be published elsewhere. The necessary material properties are mentioned solely to justify the values of the model parameters and link them to the ongoing development of the refrigerator.

In this manuscript, we use numerical simulations to compare two fundamentally different design approaches to the construction of the CNDR. Namely, the already discussed series configuration~\cite{Fukuyama,CNDR} is contrasted to the parallel configuration of the two demagnetization stages. Design considerations and the thermal models for both configurations are presented and, subsequently, numerical simulations similar to the calculations in Ref.~\cite{CNDR} are used to provide quantitative comparison of the performance of the CNDR in these two configurations, with special attention devoted to the quality of the thermal links used.

\section{Comparing CNDR and CADR}
\label{sec:CNDRCADR}

As the CADR setup is the starting point for the development of CNDR, it might be useful at this point to highlight some of the practical differences between CADR and CNDR setups due to the different operating range of temperatures. Besides the choice of the nuclear refrigerant and heat switches mentioned above, additional differences arise due to the significantly suppressed thermal conductivities of all materials at mK or sub-mK temperatures. Electronic thermal conductivity in metals is directly proportional to temperature, and hence will be reduced by three orders of magnitude between 1~K and 1~mK. Furthermore, the thermal conductivity of the heat switches in the superconducting state is suppressed by an even greater factor upon cooling, with a switching ratio as great as 10$^6$ observed at 50~mK \cite{Pobell}.

This leads to several important differences on the practical level. While serial CADR setups are often limited by the high heat leak through the final heat switch in the superconducting state (off-state), for CNDR, the opposite is true. While the off-state conductivities are negligibly low, the heat switches, along with their leads and contacts represent a significant thermal resistance even in the normal state (on-state), limiting the heat transfer rate through any such element. As a result, for the serial CNDR, the heat transfer rate between the two demagnetization stages may become comparable to the external heat leak to the sample space (both typically in the nW range), which further affects the overall performance of the refrigerator and to some extent complicates its analysis.

\section{CNDR Design and Operation}
\label{sec:1}

\begin{figure}[tb]
\includegraphics[width=0.99\linewidth]{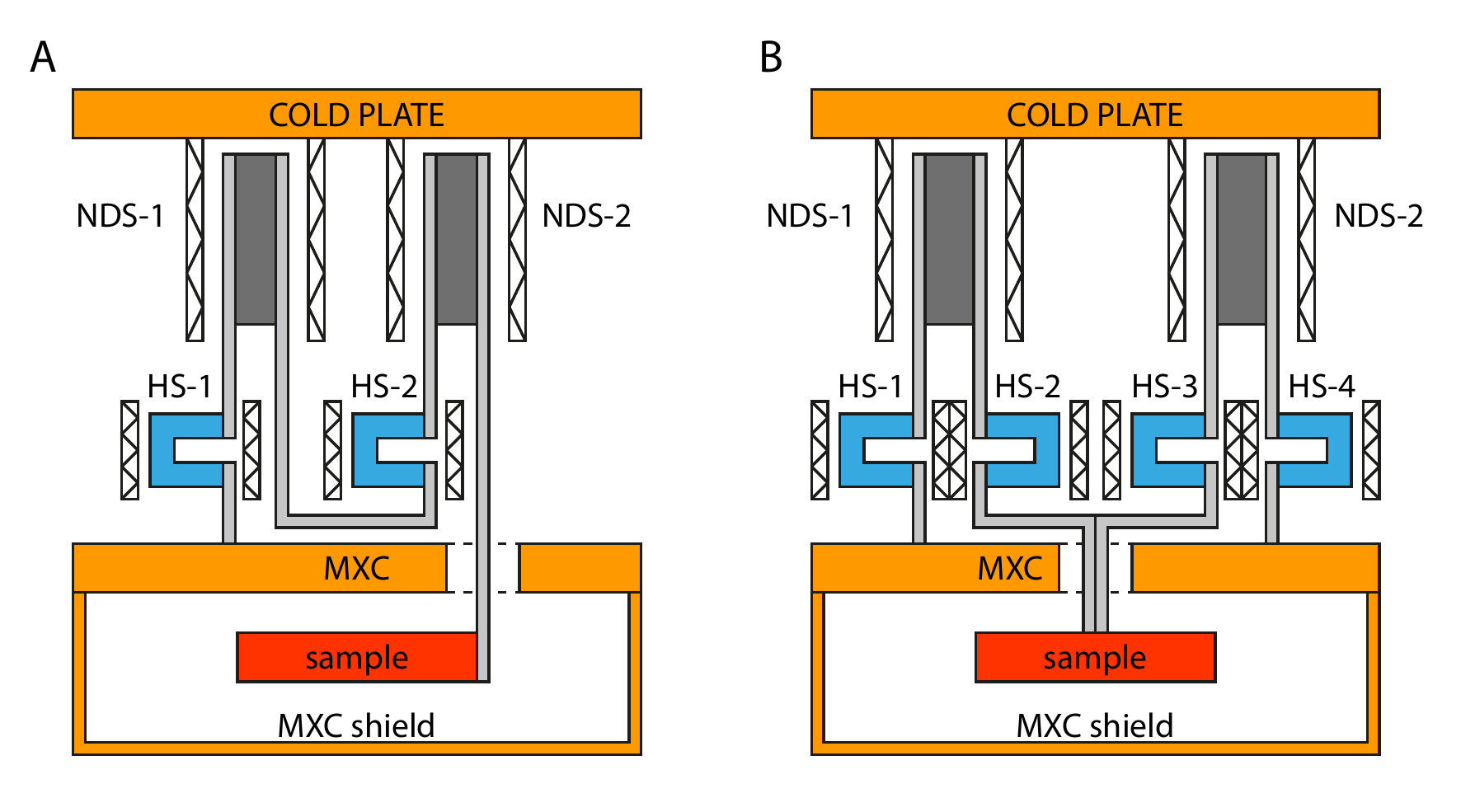}
\caption{The schematic design of serial (A) and parallel (B) configuration of the CNDR.}
\label{fig:design}       
\end{figure}

Both principal design configurations have already been discussed for CADR in Ref.~\cite{Shirron2} and are illustrated for CNDR in Fig.~\ref{fig:design}. In the series configuration (S-CNDR), the first of the nuclear demagnetization stages (NDS-1) is connected to the mixing chamber plate (MXC) of a dilution refrigerator via a superconducting heat switch (HS-1). The second stage (NDS-2) is directly linked to the sample space (load) and is separated from the first NDS via another heat switch unit (HS-2). The stages may be designed asymmetrically, with NDS-2 containing a lower quantity of the nuclear refrigerant than NDS-1, or with a different nuclear refrigerant altogether.

\begin{figure}[tb]
\includegraphics[width=0.9\linewidth]{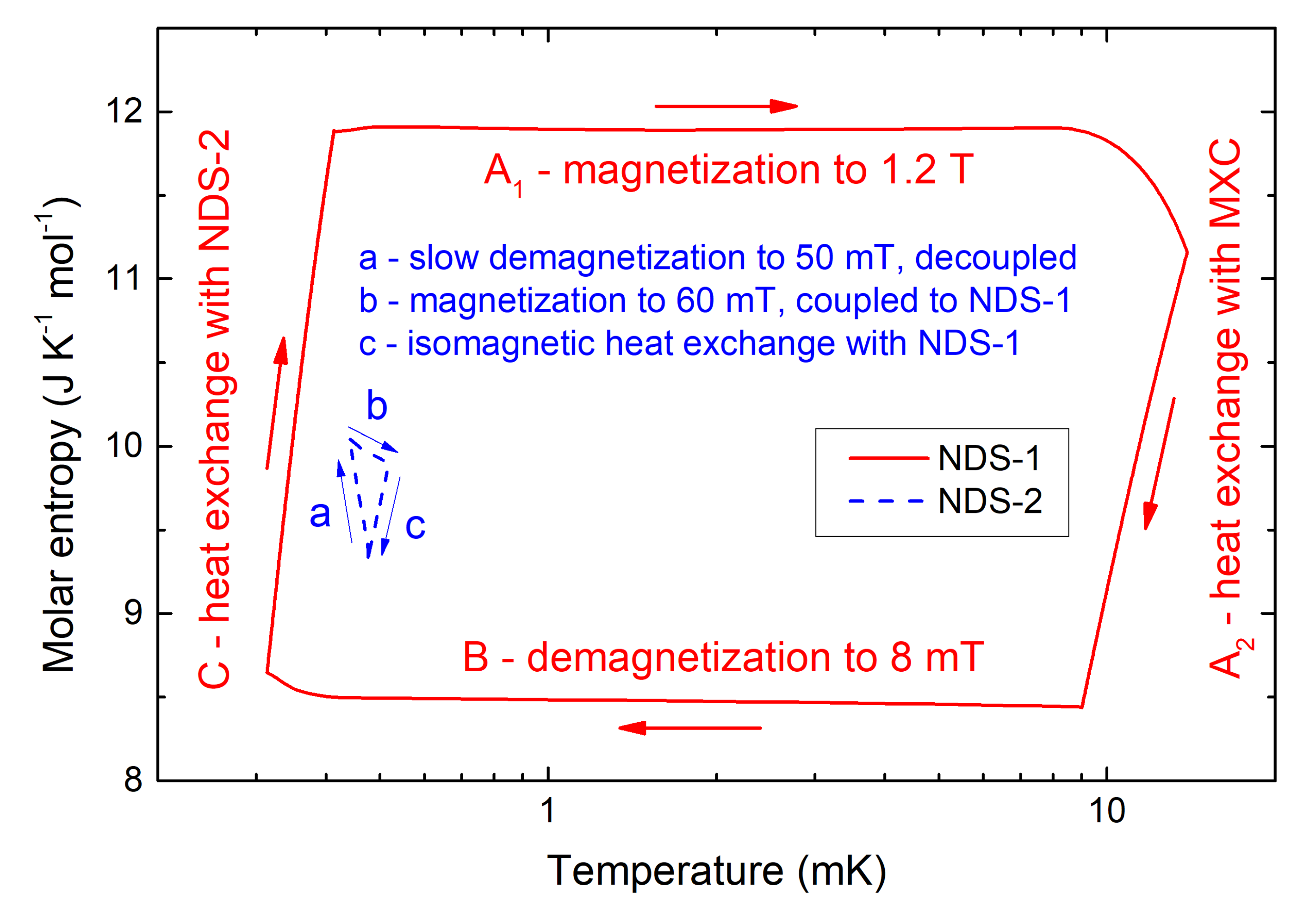}
\caption{The steady-state operation cycle of the series CNDR. In this design, NDS-1 provides precooling for NDS-2 during its up-magnetization, and rejects entropy at the MXC temperature. Ideally, the loop of NDS-2 should be as narrow as possible to minimize losses and provide a stable temperature, but with realistic durations of each step, a finite span of temperatures is necessary. The span of entropies covered by NDS-2 is determined by the choice of its maximum and minimum magnetic field values.}
\label{fig:operationS}       
\end{figure}

The operation of the S-CNDR consists of \emph{three main steps} and can be described as follows (see Fig.~\ref{fig:operationS}). Assuming steady-state operation, we start with NDS-2 at the sample temperature and at a moderate magnetic field (of order 60~mT), and thermally decoupled from NDS-1, which is at low temperature and minimum magnetic field ($\approx$10~mT). \emph{Step 1:} a slow demagnetization is started on NDS-2 (Fig.~\ref{fig:operationS}:a) to keep the sample temperature stable and offset any external heat leaks, while NDS-1 is magnetized to full magnetic field (of order 1~T) and warms up (Fig.~\ref{fig:operationS}:A$_1$). Eventually, it is thermally coupled to the MXC and allowed to exchange heat (Fig.~\ref{fig:operationS}:A$_2$), transferring its excess entropy to the MXC. \emph{Step 2:} once the final magnetic field and a temperature sufficiently close to that of the MXC are reached, NDS-1 is decoupled from the MXC and its demagnetization begins (Fig.~\ref{fig:operationS}:B), aiming for a temperature just below that of the sample space. \emph{Step 3:} once the temperature of NDS-1 drops below that of NDS-2 (this should coincide with the end of the slow demagnetization of NDS-2), the two stages are thermally linked via the HS and heat exchange between them takes place (Fig.~\ref{fig:operationS}:C), leading to transfer of entropy from NDS-2 to NDS-1. During this time, the demagnetization of NDS-1 may continue for some time until the minimum field ($\approx$10~mT) is reached, while NDS-2 undergoes a magnetization to its starting value of the magnetic field (Fig.~\ref{fig:operationS}:b) accompanied with a thermal relaxation to NDS-1 temperature (Fig.~\ref{fig:operationS}:c). We note that the heat transfer rate needed at this point is given jointly by the amount of heat produced in NDS-2 during its magnetization and the sample space heat leak. As the performance-limiting factor of the S-CNDR is the heat transfer rate between the two stages, it might be advantageous to perform the magnetization of NDS-2 in the initial phase of this step, and thus increase the total heat transferred by inducing a higher temperature difference between NDS-2 and NDS-1, albeit at the cost of reduced temperature stability of the sample space.

\begin{figure}[tb]
\includegraphics[width=0.9\linewidth]{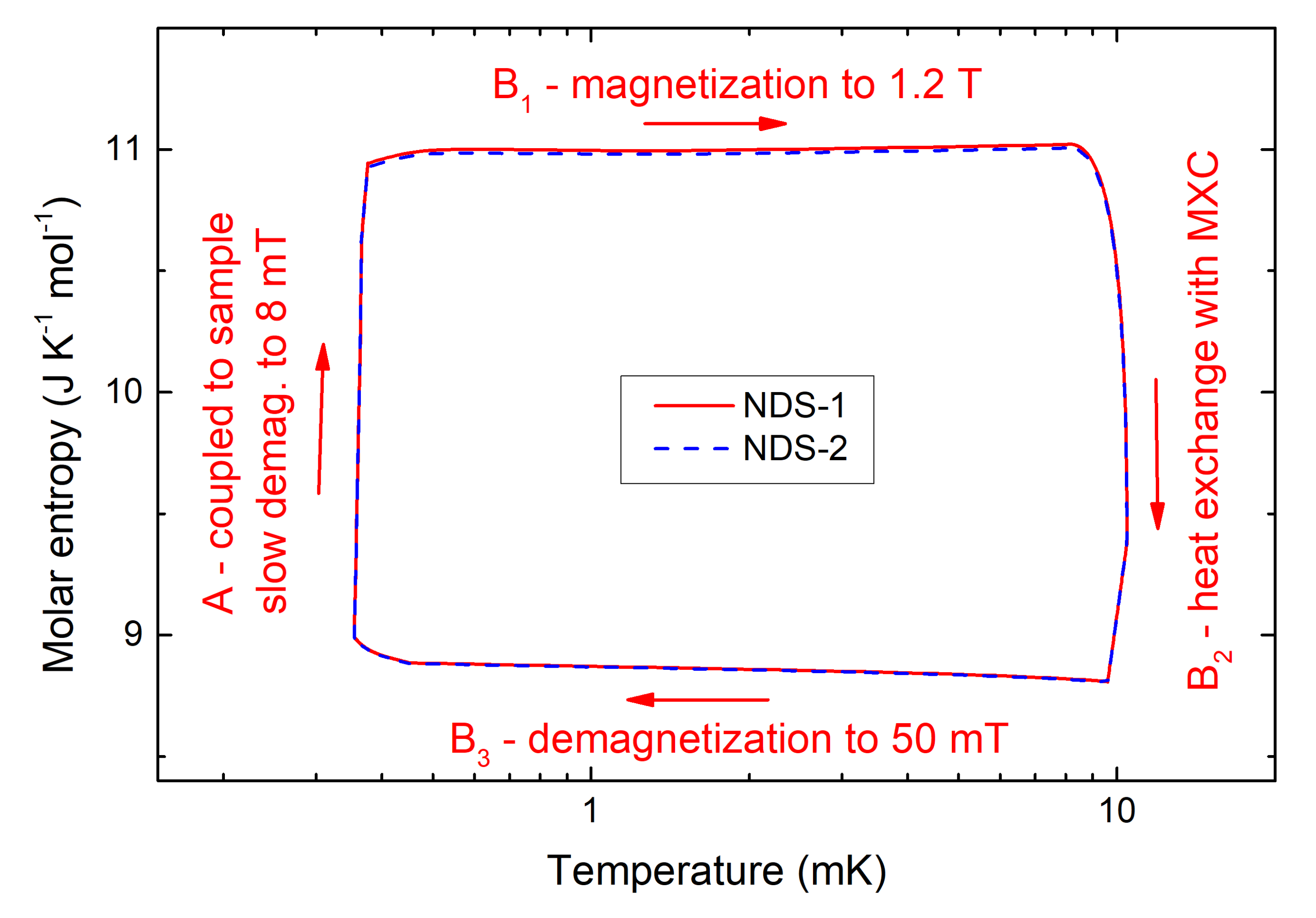}
\caption{The steady-state operation cycle of the parallel CNDR. While the initial cooldown and the first few cycles would differ slightly for both demagnetization stages, once a steady-state is reached, NDS-1 and NDS-2 follow the same operation cycle (out of phase with each other) and both reject entropy at the MXC temperature. As the rate of entropy rejection per cycle is effectively doubled compared to S-CNDR, a higher cooling power is expected for the P-CNDR configuration.}
\label{fig:operationP}       
\end{figure}

On the other hand, the parallel configuration (P-CNDR) requires each of the two NDS to be linked to both the MXC and the sample space via separate HS units. In this case, both NDS should be designed symmetrically, with balanced heat capacities. The extra heat switches necessary for the P-CNDR configuration represent some design complications as each HS unit requires its own small superconducting solenoid which must be mounted on the dilution refrigerator and operated independently. When choosing between S-CNDR and P-CNDR, one must therefore weigh any benefits found in the performance of the P-CNDR against the increased complexity of the setup and its larger footprint on the dilution refrigerator.

The operation of the P-CNDR has \emph{two main steps} (see Fig.~\ref{fig:operationP}) during which either of the NDS is in thermal contact with the sample space and providing the cooling via a slow demagnetization (Fig.~\ref{fig:operationP}:A), while the other NDS is being ``regenerated'' at the MXC temperature. The regeneration procedure involves decoupling from the sample and magnetization to the maximum field (Fig.~\ref{fig:operationP}:B$_1$), heat exchange with the MXC (Fig.~\ref{fig:operationP}:B$_2$), followed by thermal decoupling and demagnetization to or below the sample temperature (Fig.~\ref{fig:operationP}:B$_3$), where the newly regenerated NDS will be switched for its counterpart. We note that the P-CNDR configuration to some extent mitigates the requirements on the quality of the thermal links, as the necessary heat transfer rate through HS-2 or HS-3 is given only by the sample space heat leak, whereas for S-CNDR, the heat produced by the magnetization of the second stage needs to be transferred through HS-2 as well.

Based on the comparison with CADR setups~\cite{Shirron2}, one would expect a higher cooling power for the P-CNDR setup than for S-CNDR, as with the two stages in parallel, heat rejection at MXC occurs during a longer part of the operating cycle. Nevertheless, it remains to be seen if other complications, specific to ultra-low temperatures arise due to the increased number of heat switches and thermal links, potentially influencing the overall heat leak in the sample space or introducing additional dissipative processes into the operation of the refrigerator, such as heat switch manipulation.

\section{Thermal Models}
\label{sec:2}

\begin{figure}[tb]
\includegraphics[width=0.99\linewidth]{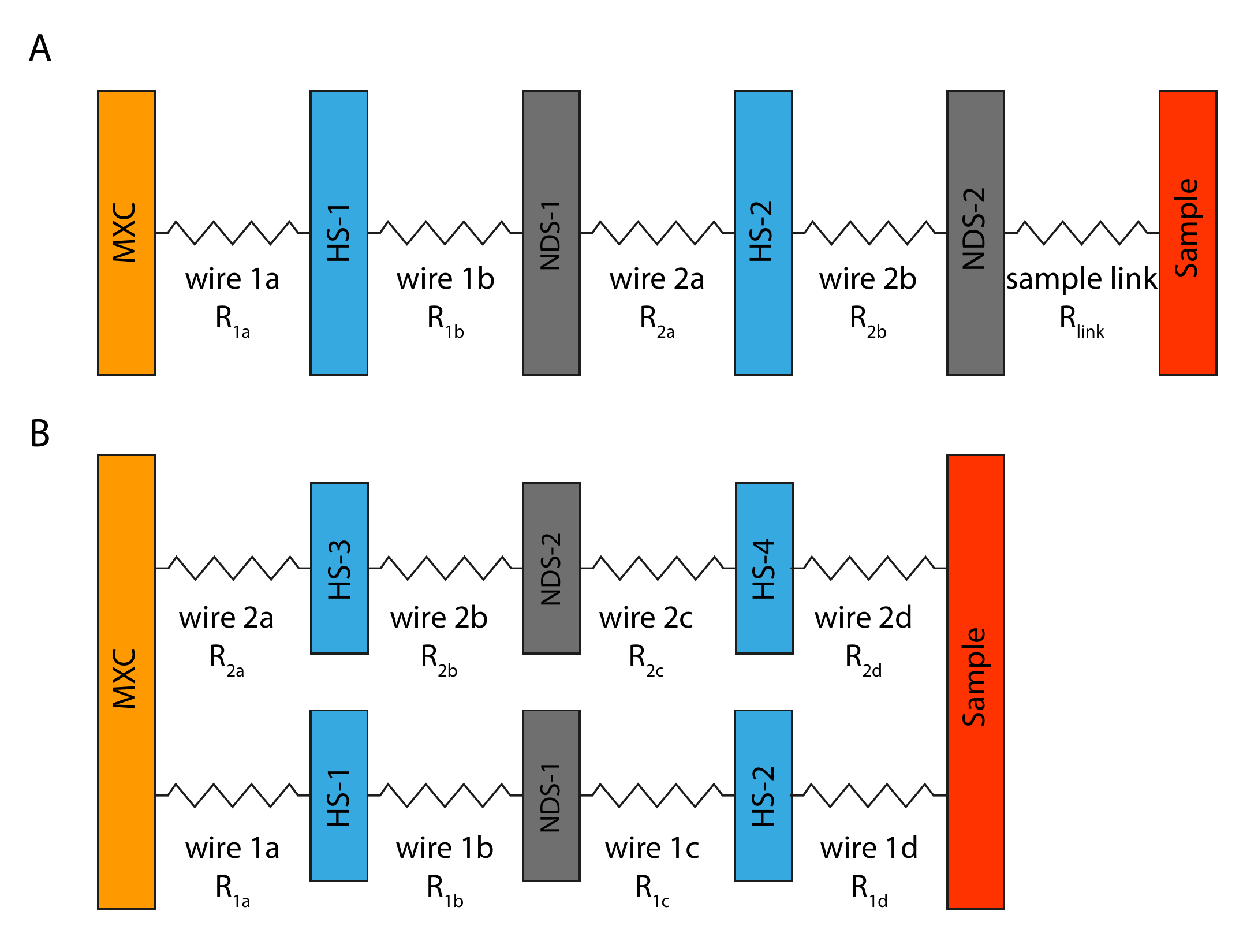}
\caption{Thermal models of series (A) and parallel (B) CNDR setups. Note that in the discussion of the serial CNDR setup in Ref.~\cite{CNDR}, the sample link is assumed to have negligible thermal resistance. This can be partly justified by the final design of this element, if a larger piece of high-conductivity metal (copper or silver) is used to cover most of the distance from the stage to the sample space. However, thinner wires would still have to be used to connect to the PrNi$_5$ inside its solenoid. While this would not greatly impact the results in Ref.~\cite{CNDR}, as NDS-2 is permanently connected to the sample space and has a much higher heat capacity, the results nonetheless represent an ideal case. On the other hand, in the simulations of P-CNDR discussed here, no such idealized thermal links are used.}
\label{fig:thermal}       
\end{figure}

The thermal models corresponding to both configurations are shown in Fig.~\ref{fig:thermal}. In the S-CNDR setup, the thermal link between NDS-1 and NDS-2 (including the heat switch HS-2) is the crucial component of the entire setup~\cite{Fukuyama,CNDR}, as it determines the heat exchange rate at which NDS-2 can be precooled from NDS-1 during its magnetization. The duration of this step, representing the longest part of the refrigerator cycle, together with the heat leaks involved determine the efficiency of the S-CNDR setup and hence the final temperature it can attain.

The necessary heat exchange between NDS-1 and NDS-2 could be, in principle, shortened if a lower capacity stage was used for NDS-2, or if the range of applied magnetic fields is reduced. Depending on the intended application and the required stability of sample space temperature, it may be possible tune the heat capacity of the second stage, together with the duration of the refrigerator cycle. An optimum choice exists, leading to peak performance. For the purposes of a direct comparison with P-CNDR, we assume a symmetrically designed S-CNDR system, with the ``active'' heat capacity of the second stage determined by the limiting values of applied magnetic field, even though this need not be the optimum case in terms of S-CNDR performance.

In the P-CNDR setup, properties of the thermal links between the NDS stages and the sample space (Fig.\ref{fig:thermal}: wires 1d and 2d) determine the rate of maximum available cooling power at a given temperature. As mentioned in the previous Section, less heat needs to be transferred at low temperature via these links than for the S-CNDR setup and hence the requirements on the equivalent electrical resistance of the thermal links may be expected to be less stringent. Nevertheless, thermal decoupling due to a finite thermal resistance between the sample and the stages will necessarily reduce the refrigeration efficiency, as any heat removed from the sample will generate additional entropy, in excess of the amount that would be generated if the stage and the sample were at the same temperature. This is fully taken into account in the numerical simulations presented below.

To directly compare the two configurations, we consider in each case two identical demagnetization stages, each containing 0.2~mol of PrNi$_5$. The thermal behaviour of the linking elements (wires) is modeled based on an equivalent electrical resistance from the Wiedemann-Franz law, assuming the wires would be made from a high-purity metal such as silver or copper. The same electrical resistance is then used for all the wires.

For most simulations, the heat capacity of the sample space is chosen so to approximate that of 10~g of high-purity Cu at 1~mK and zero magnetic field. Specific heat capacity of 1.15$\times 10^{-5}$~J kg$^{-1}$ is used, in close agreement with the electronic contribution to the heat capacity of Cu given in Ref.~\cite{Pobell}). We note that with the same heat load and thermal resistance, the simulation results do not change appreciably if a mass of 100~g or 1000~g is used for the sample space instead. Additional simulations were performed with a sample configuration modeling 100~g of Cu in a magnetic field of 100~mT, including the much larger heat capacity of the nuclear magnetic moments~\cite{Pobell}. As expected, the extra heat capacity had no effect on the final temperatures obtained, but resulted in a longer settling time, in this case comparable to the duration of one cycle of the CNDR.

The thermal models of individual components of the setup are detailed in Ref.~\cite{CNDR}. The thermal resistance, $R(T)$, of any linking element consisting of an Al heat switch and connecting leads is given as the sum of the thermal resistances of all its parts and of boundary (contact) resistances. For the Al heat switches as such, different relations were employed for their normal and superconducting state, describing heat conduction by electrons ($\propto T$) and phonons ($\propto T^3$). A residual resistivity ratio (RRR) of 11000 was used to describe the normal state, resulting in heat conductivity $\kappa_{\rm N} = 1.01 \times 10^4 \rm{[W m^{-1} K^{-2}]} T$, while the heat conductivity $\kappa_{\rm S} = 0.68 \rm{[W m^{-1} K^{-4}]} T^3$ was used for the superconducting state, yielding a switching ratio $\kappa_{\rm S} / \kappa_{\rm N} = 6.73 \rm{[K^{-2}]} T^2$, in agreement with the low temperature limit of Al heat switch properties given in Ref.~\cite{Pobell}. For convenience, the contact thermal resistances were modeled using a temperature-independent equivalent electrical resistance and the Wiedemann-Franz law. A similar approach has been used to model the wires in thermal links. For the nuclear refrigerant, an empirical model of its thermo-magnetic properties was used~\cite{CNDR} that interpolates between the paramagnetic behaviour at higher temperatures and the ferromagnetic ordering that sets in near 0.4~mK. Agreement with experimental data of Ref.~\cite{Kubota} was the driving criterion in the devising of this model (see Ref.~\cite{CNDR}).

Additional heat leaks (vibrational heating $\dot{Q}_{vibr} = 10^{-8} \rm{[W T^{-1}]} \left|B\right|$, eddy current heating $\dot{Q}_{eddy} = 0.03 \rm{[W T^{-2} s^2]} \dot{B}^2$, plus an additional constant heat leak of 2~nW) are considered for each nuclear demagnetization stage in the same way as in Ref.~\cite{CNDR}. The dependencies used and the values of the prefactors are adapted from Ref.~\cite{Parpia}, where such behaviour was observed experimentally on a nuclear demagnetization refrigerator in the relevant range of magnetic fields. We note that other works have reported different dependencies of the vibrational heating on the magnetic field~\cite{Todoshchenko}, and that in general, vibrational heating will depend on the details of the experimental setup. The numerical computations are again described in detail in Ref.~\cite{CNDR}, as well as the model of entropy of PrNi$_5$ used throughout this work.

\section{Results and Discussion}
\label{sec:3}

For the comparison of the performance of the two CNDR configurations, the data on final temperature vs. heat leak from Ref.~\cite{CNDR} will be used for the S-CNDR setup, while new results will be presented here for the P-CNDR setup.

\begin{figure}[tb]
\centering
\includegraphics[width=0.95\linewidth]{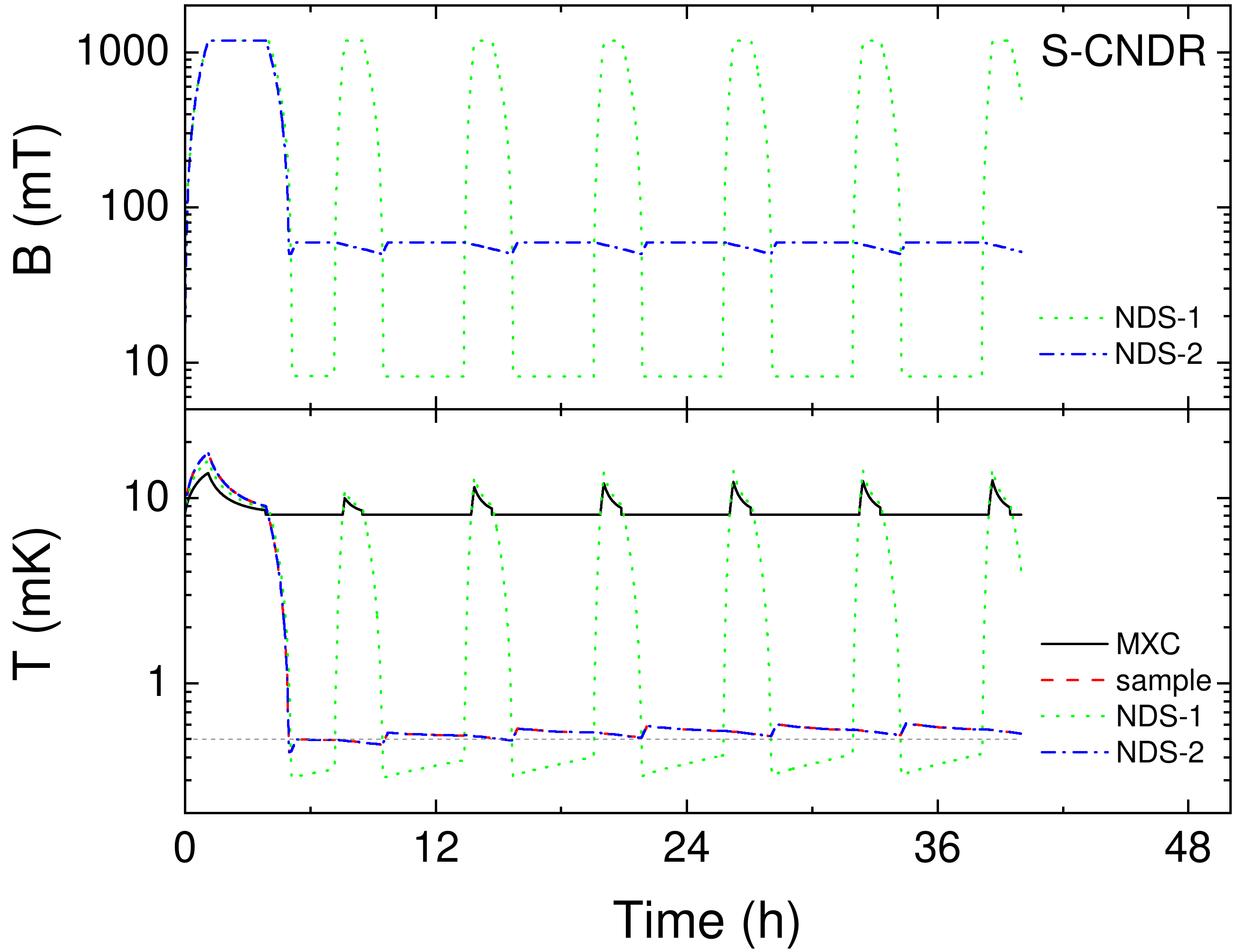}
\includegraphics[width=0.95\linewidth]{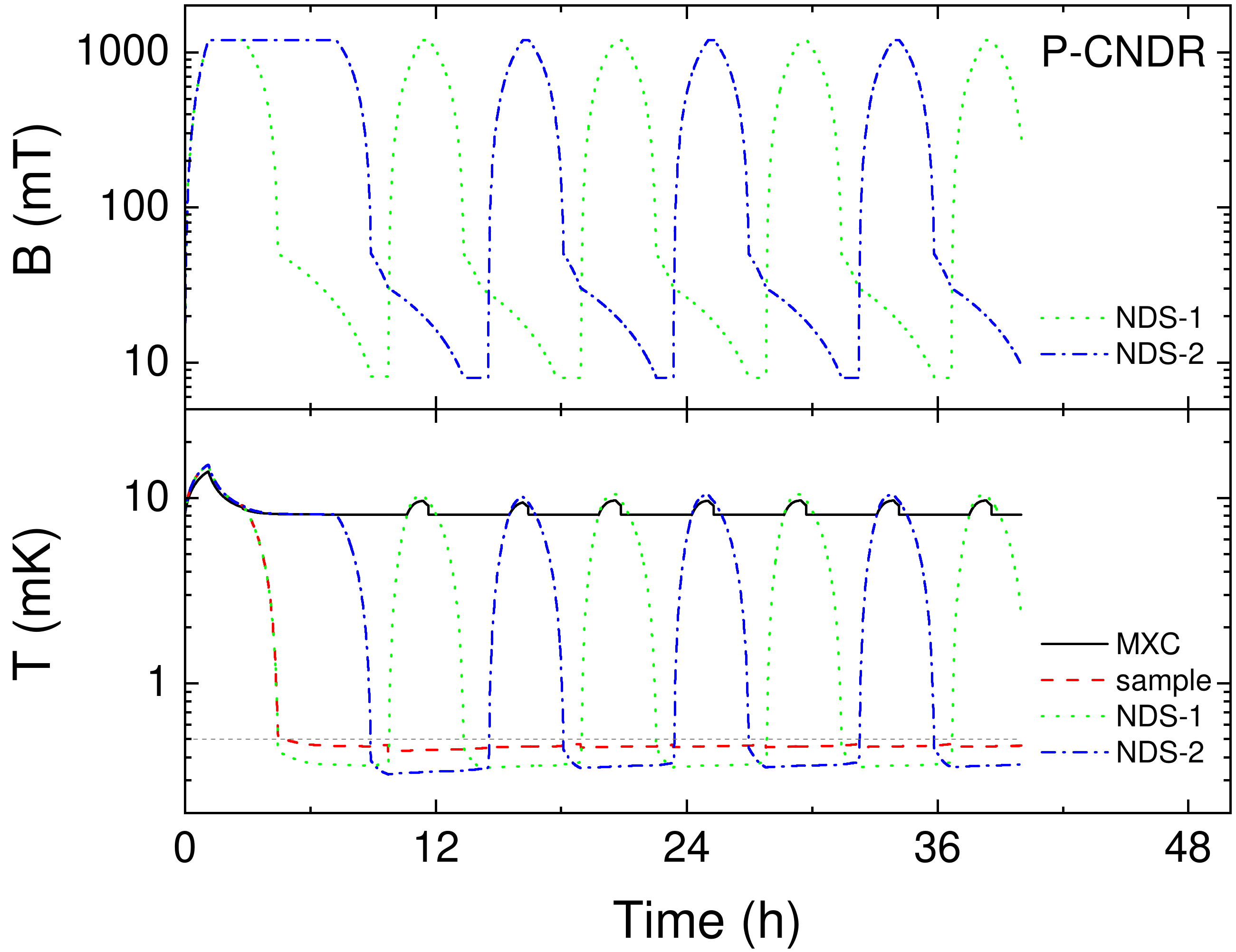}
\caption{Time traces of magnetic fields and temperatures for the S-CNDR (top) and P-CNDR (bottom) setups, as given by the numerical simulations for a 150~n${\rm \Omega}$ equivalent electrical resistance of the thermal links and 5~nW sample space heat leak. Parabolic profiles for magnetization/demagnetization are used. After a stabilization time of approximately two cycles, a steady state is reached and the maximum sample temperature during a cycle can be extracted. The durations of (de-)magnetization steps and values of magnetic fields are tuned by hand to achieve near-optimum performance. The starting conditions are chosen with the MXC temperature of $\approx$8~mK and the maximum magnetic field used on the demagnetization stages is 1.2~T, as in Ref.~\cite{CNDR}. The (gray) dashed line marks the temperature of 0.5~mK.}
\label{fig:timetraces}       
\end{figure}

First, a direct comparison of sample time-traces of temperatures and magnetic fields for the S-CNDR and P-CNDR setups is given in Fig.~\ref{fig:timetraces} for a 150~n${\rm \Omega}$ equivalent electrical resistance of the thermal links (modeling two bundles of copper wires and an aluminium heat switch in normal state) and 5~nW sample space heat leak (representing an optimistic estimate, see, e.g., Ref.~\cite{Parpia} for comparison). The small jumps in sample temperature for the S-CNDR setup are due to the relatively rapid magnetization of the second stage. While these could be avoided by magnetizing more slowly, the overall performance would be degraded due to a lower net heat transfer between the stages, resulting in a higher cycle-averaged temperature. Regardless, the S-CNDR setup fails to reach the 0.5~mK mark (highlighted in Fig.~\ref{fig:timetraces}) even in the cycle minimum temperature, while the P-CNDR setup, once cooled down, stays below this mark consistently.

\begin{figure}[tb]
\includegraphics[width=0.99\linewidth]{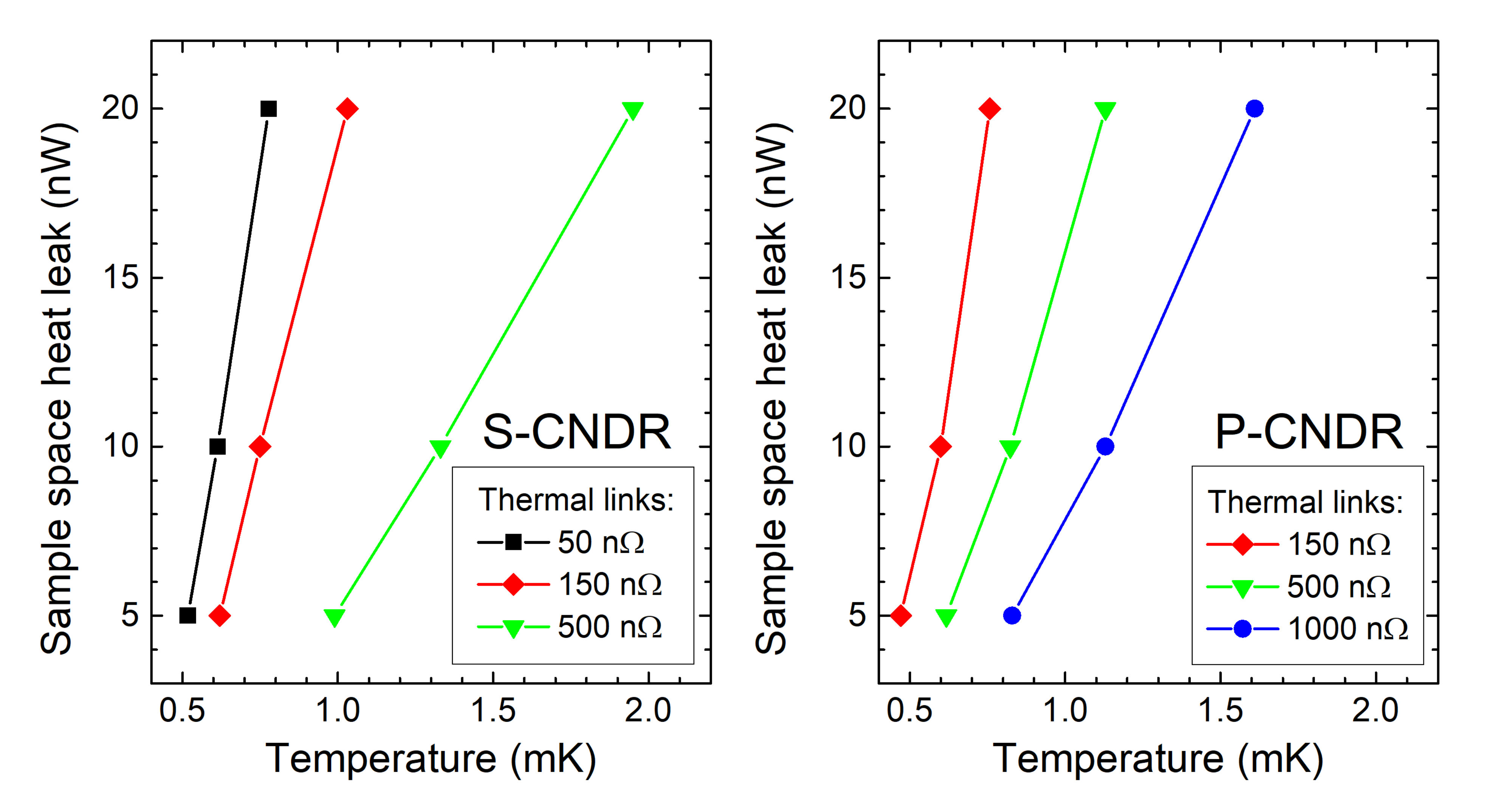}
\caption{Comparison of the final temperature (maximum sample space temperature during a cycle) vs. sample space heat leak for the serial (left) and parallel (right) CNDR setups. The data in the left panel are adapted from Ref.~\cite{CNDR}. It is clearly shown that the P-CNDR setup can achieve lower temperatures with the same quality of thermal links under the same heat load. Conversely, this implies that at a given temperature (in the mK range), the P-CNDR setup indeed has a higher cycle-averaged cooling power than the S-CNDR setup, as expected. We note that if cycle-averaged temperatures were used instead of the maximal ones, this conclusion would hold, see text and Table~\ref{tab:temperatures}.}
\label{fig:comparison}       
\end{figure}

Similar numerical simulations as those shown in Fig.~\ref{fig:timetraces} were performed for the P-CNDR setup operating with different values of the equivalent electrical resistance of the thermal links and different values of sample space heat leak. For each case, steady-state operation was established and the maximum sample temperature within a cycle was obtained. These results are summarized in Fig.~\ref{fig:comparison} and compared to the data of Ref.~\cite{CNDR}. For the S-CNDR setup, due to considerable temperature variations present, cycle-averaged temperatures were obtained as well. All values of temperature are listed in Table~\ref{tab:temperatures}. In all cases, the P-CNDR setup reaches lower temperatures than the S-CNDR setup.

\begin{table}
	\centering
		\begin{tabular}{c|c|c|c|c}
			Link resistance & Heat leak & S-CNDR max. & S-CNDR avg. & P-CNDR max. \\
			(n$\rm{\Omega}$) & (nW) & (mK) & (mK) & (mK)\\
			\hline
			50 & 5 & 0.52 & 0.48 & --\\
			50 & 10 & 0.61 & 0.57 & --\\
			50 & 20 & 0.78 & 0.70 & --\\
			150 & 5 & 0.62 & 0.57 & 0.47\\
			150 & 10 & 0.75 & 0.70 & 0.60\\
			150 & 20 & 1.03 & 0.94 & 0.76\\
			500 & 5 & 0.99 & 0.94 & 0.62\\
			500 & 10 & 1.33 & 1.29 & 0.82\\
			500 & 20 & 1.95 & 1.90 & 1.13\\
			1000 & 5 & -- & -- & 0.83\\
			1000 & 10 & -- & -- & 1.13\\
			1000 & 20 & -- & -- & 1.61
		\end{tabular}
	\caption{Final temperatures obtained from numerical simulations of the S-CNDR and P-CNDR setups. For the S-CNDR setup, both maximal and cycle-averaged temperatures are shown, for P-CNDR, only the maximal temperatures are given. See also Fig.~\ref{fig:comparison}.}
	\label{tab:temperatures}
\end{table}

Moreover, the results indicate that the rather stringent requirements for the operation of the S-CNDR setup (thermal links with equivalent electrical resistance comparable to 150~n${\rm \Omega}$) are to some extent alleviated in the P-CNDR configuration. Temperatures close to 1~mK should be attainable with P-CNDR even with significantly worse thermal links of 500~n${\rm \Omega}$ equivalent resistance under a heat load of 20~nW.

\section{Conclusions}
\label{sec:4}
While one must bear in mind that the simulations do not necessarily describe all experimental facts accurately, our results clearly show that, given the same quality of thermal links, the P-CNDR configuration has superior performance to the S-CNDR design in terms of the final temperature reached at a given heat leak, or in terms of the cooling power attained at a given temperature.

We note, however, that the superior cooling power of the P-CNDR comes at the cost of additional system complexity. Additionally, this design may lead to reduced stability of sample space temperature not captured in the presented simulations, as switching the sample connection between the two stages may in practice induce significant temperature variations due to the required manipulation of two heat switches. Indeed, one drawback of P-CNDR is the necessity to operate a total of four heat switches together with their solenoids instead of just two such units in the S-CNDR design, leading to additional heat leaks and delays within the refrigerator cycle along with the setup footprint and complexity. Nevertheless, aside from special cases that may be better suited to the S-CNDR design, we expect the P-CNDR to perform better than the S-CNDR for a given thermal link conductance, and the P-CNDR design clearly warrants experimental study.

\begin{acknowledgements}
We acknowledge support from the ERC StG grant UNIGLASS No. 714692 and ERC CoG grant ULT-NEMS No. 647917. The research leading to these results has received funding from the European Union’s Horizon 2020 Research and Innovation Programme, under Grant Agreement no 824109. DS acknowlegdes support from the Czech Science Foundation (GACR) under project No. 20-13001Y.
\end{acknowledgements}



\end{document}